\documentclass{cjaa}
\usepackage{graphicx}                   %for PS/EPS graphics inclusion, new
\input{epsf.sty}                        %for PS/EPS graphics inclusion, old
\input{psfig.sty}                       %for PS/EPS graphics inclusion, old

\def\be{\begin{equation}}
\def\ee{\end{equation}}
\def\bea{\begin{eqnarray}}
\def\eea{\end{eqnarray}}
\def\nn{\nonumber}

\def\s2p{\sqrt{2\pi}}
\def\rf{r_{\rm F}}
\def\rdiff{r_{\rm diff}}

\def\rref{r_{\rm ref}}
\def\rft{r_{\rm F}^{2}}

\def\p2{\phi''}
\def\sp2{\sigma_{\phi''}}

\def\L0{L_{0}}
\def\la{\langle}
\def\ra{\rangle}
\def\uj{u_{j}}

\begin{document}
 %  \date{Received~~2005 month day; accepted~~2005~~month day}

\title{The stationary phase point method for transitional
scattering: diffractive radio  scintillation}

\author{C.M. Zhang \inst{}\mailto{}  }
   \offprints{C.M. Zhang}
   \institute{
  National Astronomical Observatories,
    Chinese Academy of Sciences,
    Beijing, China
   \email{zhangcm@bao.ac.cn} }

\date{\today}
\abstract{ The stationary phase point (SPP)  method in
one-dimensional case
  is introduced  to
treat the diffractive scintillation.  From weak scattering, where
the SPP number  N=1,  to strong scattering (N$\gg$1),  via
transitional scattering regime (N$\sim$2,3),  we find that  the
modulation index of intensity experiences  the monotonically
increasing from 0 to 1 with
    the scattering strength,  characterized by the ratio of Fresnel
scale $\rf$ to   diffractive scale $\rdiff$.
%
% Applications of SPP method  to the   radio scintillation
%observations of  pulsars and quasars--IDVs are mentioned.
 {\keywords  Radio scintillations --- pulsar general
  --- interstellar medium }
}
   \authorrunning{C.M. Zhang}
   \titlerunning{SPP method for transitional scattering }

\maketitle

\section{Introduction}
As the radio  waves  propagate in  the interstellar medium
(ISM), %where the inhomogeneities  arises the scintillations,
 the diffraction and   refraction introduced by small-scale
 and large-scale inhomogeneities  lead  to  the  flux variations
 or scintillations (review see Rickett, 1977, 1990, 2006; Narayan 1992).
The  radio wave  interference fringes are seen   evolving  with
time, due to the motions of source, observer and ISM, which   can be
explained in terms of wave scattering  through a random medium, with
the electron density inhomogeneities being described by a power-law
spectrum, close to the Kolmogorov spectrum (Armstrong, Rickett and
Spangler 1995; Cordes, Weisberg and Boriakoff 1985).
As for the characteristic scales responsible for the
 scintillations,  Fresnel length scale is defined by
 $r_{\rm F}=\sqrt{z/k}$, where   $z$ is the distance
  between the scattering screen and the
observer's plane and $k=2 \pi/\lambda$ is the wave number, and the
diffractive scale $r_{\rm diff}$ is defined by writing the phase
structure function in the forms $D(r)=(r/r_{\rm diff})^{\alpha}$ for
a Kolmogorov spectrum of turbulence
(see e.g. Narayan 1992 for a  review).
 In the weak scattering regime,  $\rf/\rdiff<<1$,
the flux variation  is   interpreted in terms of weak focussing due
to phase curvature on the scattering screen on scale  $\sim r_{\rm
F}$. Whereas, in the strong scattering regime ($\rf/\rdiff>>1$)
there are two variation scales, small  diffractive scale  $\rdiff$
and large refractive scale  $\rref=\rf^{2}/\rdiff$,   caused by
interference between the many coherent patches of size $r_{\rm
diff}$ over the scattering screen.
However, in  the transitional scattering regime, $\rf/\rdiff\sim 1$,
 which  is  applicable to most intra-day variable
extragalactic radio sources observed at frequencies between 1 and
10~GHz (see, e.g. Walker \& Wardle 1998),  the flux variation scale is a
little fuzzy because $\rf$, $\rdiff$ and $\rref$ are very similar,
 perhaps a  mixing variation scale may be produced.
The two time scales of spectrum intensity from the observations of
pulsars reveal variations of minutes to hours  as a diffractive
 scintillation  (see e.g. Scheuer
1968;  Manchester and  Taylor 1977; Lyne \& Smith 2005) and days to
months as a refractive  scintillation (see, e.g. Rickett 1984
 for reviews;  Kaspi \& Stinebring 1992; Wang et al. 2005).

%
%The diffractive  scintillations  have been approximately  described
%by the modulation index m of the spectrum intensity, $m<<1$ and
%$m\sim 1$ in the weak and strong scattering regimes respectively
%(see e.g. Narayan 1992).
%
In  the transitional scattering regime,
 where both geometric optics and wave optics are equally important,
   with $r_{\rm diff}$ comparable to $r_{\rm F}$, the  structures on
the scattering screen have a focal length comparable to the distance
from the scattering screen, so the role of the  caustic focussing
will effect here.

The  paper  briefly introduces the  unified descriptions
of modulation index from weak, transitional and strong scattering by
 the stationary phase point  method (SPPM),
which  has been previously
 paid attention by  Gwinn et al. (1998) and recently applied  to
interpret the parabola phenomena occurring in the pulsar secondary
spectrum (Walker et al. 2004).
%
%Therefore, the  paper is organized as follows. We   begin   the
%stationary phase approximation  in \S2,   and the calculations and
%conclusions  on the modulation index  are  presented in \S3.

\begin{figure}
\vspace{50mm}
   \begin{center}
\includegraphics{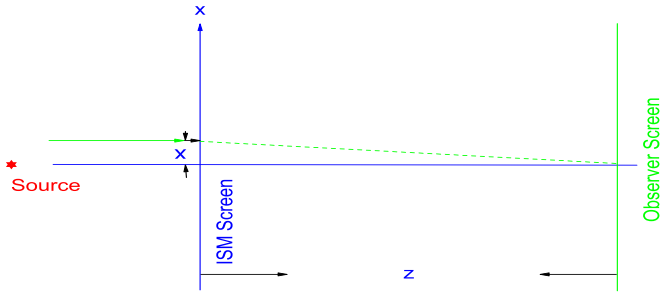}
%\vspace{60mm}
\caption{ The illustration of one dimensional scattering screen.}
\label{psr-ism-scin-simp}
   \end{center}
   \end{figure}

%  are  important in the strong scattering regime.
%Geometric optics suffices to account for the properties of
%refractive scintillation  (e.g. Blandford \& Narayan 1985).  Flux
%variations are attributed to the focussing and defocussing caused by
%phase curvature on the scattering screen on scales of size $\sim
%r_{\rm ref}=r_{\rm F}^2/r_{\rm diff}$.
%

\section{The stationary phase approximation}\label{Stationary}
We consider  the  wavefield from a point source at infinity to be
incident up on a single thin, one-dimensional scattering screen, as
described in figure \ref{psr-ism-scin-simp}. For an incident
wavefield of unit amplitude the wavefield observed at the central
position on the line a distance $z$ (${r_{\rm F}^{2}} = z/k$) from
the scattering screen is (e.g. Born \& Wolf 1980) \be u(z) =
\frac{1}{\sqrt{2 \pi r_{\rm F}^{2}}} \int_{-\infty}^{\infty} d{ x}
\exp\left[i \left(\frac{x^{2}}{2 r_{\rm F}^{2}} +  \phi({ x})
\right) \right].\label{uz} \ee

To solve this we employ the method of stationary phase (e.g. Mandel
\& Wolf, P.128), which states that an integral of the form,  \bea
F(k) = \int_{-\infty}^{\infty}f(x) e^{i k g(x)} \, dx, \eea has an
approximation containing contributions from critical points inside
the boundary of integration $(-\infty, \infty)$: \bea F(k) \sim
\left( \frac{2 \pi}{k} \right)^{1/2} \sum_{j=1}^{n}
\frac{\epsilon_j}{| g''(x_{j}) |^{1/2} } f(x_j) e^{i k g(x_j)}, \eea
where $ \epsilon_j = e^{\pm i \pi/4}$ \,\, is  according \,\, as
\,\, $g''(x_j) > 0 { \; } (<0) $ for + (-)   and $x_j$ are the
points of stationary phase which satisfy the condition {\bf $
g'(x_j) = 0. $  Applying the stationary phase approximation  and
 $g(x) = k^{-1} (\phi(x) + x^{2}/2 r_{\rm
F}^{2})$  to Eq.(\ref{uz}), one has}

%\bea
%u(z) &\approx&
% \frac{1}{\sqrt{2 \pi r_{\rm F}^{2}} }
%\left\{ ( 2 \pi )^{1/2} \sum_{j=1}^{N}
%\frac{\epsilon_j}{\sqrt{|\frac{1}{r_{\rm F}^{2}} + \phi''(x_j)|} }
%\exp [ix_j^{2}/2 r_{\rm F}^2 + i \phi(x_j)] \right\}. \nn\\

\be
 u(z)  \approx \sum_{j=1}^{N} \uj\;, \;\;\;
 \uj = \frac{\epsilon_j}{\sqrt{| 1 + r_{\rm
F}^{2}\phi''(x_j)|} } \exp [ir_{\rm F}^{2} \phi'^{2}(x_j)/2 + i
\phi(x_j)]. \label{statphaseU} \ee

{\bf  It is noted that Eq.(3) or Eq.(4) is an approximate result,
and it is assumed that g''(x) does not go to zero. The independent
variables are the phasor $\phi$, its first   derivative $\phi'$, and
the total number of them is 2N,   where N is the SPP number.}
 The mean intensity of the unit
amplitude wave, $<I>=1$, is identified as the average value of the
second moment of the amplitude $<uu^{*}>$. Adding a subscript $N$ to
denote the number of SPPs, in the SPP approximation this is \be
<I>_{N} =\sum_{i,j=1}^N<u_iu^*_j>= \sum_{i=1}^N<u_i u^*_i >
+\sum_{i\ne j=1}^N<u_i u^*_j > = 1. \label{avi} \ee Thus the number
of SPP is the maximum integer   of N solved from Eq.(\ref{avi}).
While considering  a statistical model in which each SPP is assumed
to come from a distribution,  all SPPs are statistically equivalent.
Then, for example, $<u_i u^*_i >$ may be written as $<u_1u^*_1>$,
and the sum over $i=1,\ldots N$ gives $N <u_1 u^*_1>$. Only two
independent averages appear in the intensity, and we write these as
\be U_{11}=<u_1 u^*_1> ,  \;  U_{12} = (<u_1 u^*_2
>+<u_2 u^*_1>)/2, \label{mean1} \ee so that Eq.(\ref{avi}) reduces to
\be <I>_{N}=NU_{11}+N(N-1)U_{12}=1. \label{avin} \ee Similar to
 the procedure  of treating the intensity, the intensity square is
 obtained  as follows,
\bea & &<I^{2}>_{N}   =    N U_{1111} + 4N(N-1)U_{1112}
 +   N(N-1) (U_{1122} + U_{1221} + U_{1212})\nn \\
& & +  N(N-1)(N-2)(4U_{1123} + 2U_{1213})  +
(N-1)(N-2)(N-3)U_{1234}\;, \label{avi2n}\eea where the terms, such
as $U_{1111}$ and $U_{1122}$, are the averaged values of the
products of  amplitudes and generally  denoted by, \be U_{jklm}
=<jklm> = < u_j u_k^* u_l u_m^*>, \ee  which satisfies, \be
<jklm>^{*} = <kjml> = <mlkj>, \ee so we have $U_{1122} = U_{1221} =
U_{11} U_{22} = U_{11}^{2}$. Henceforth, the modulation index $m$ of
diffraction   is defined  by \be
%m =  <I^{2}>_{N}/<I>^{2}_{N} - 1 = <I^{2}>_{N} - 1
m =  \sqrt{<I^{2}>_{N} - 1}\;. \ee

\subsection{The calculations of the intensity in SPP}
The power spectrum of the phase inhomogeneities, $\Phi(k)$,
 is assumed to have the following form
 (see e.g. Blandford \& Narayan 1985; Goodman \& Narayan 1985)
\bea \Phi(k) = \left\{ \begin{array}{ll}
L_0,   & k <1/L \\
L_0 (L k)^{-\beta}, & 1/L < k < 1/l \\
0, & k>1/l
\end{array}\right. ,
\label{powers} \eea where $\beta = 8/3$ is an index for a Kolmogorov spectrum
of turbulence with the structure constant $L_{0}$
and  $l$ ($L$) is  the inner (outer) scale.
 In addition, we assume the phase parameters to follow
the Gaussian distribution $P(\psi)$, where $\psi$ represents the
phase variables $\phi$, $\phi'$ and $\phi''$, \bea
    P(\psi) =  \frac{1}{\sqrt{2 \pi \sigma_{\psi}^2}} \exp
    \left[- \frac{{\psi}^2}{2 \sigma_{\psi}^2} \right],
    \label{gauss}
\eea where $\sigma_{\psi}$ is a standard deviation of the parameter
 $\psi$,
 calculated from the spectrum function of Eq.(\ref{powers})
 (see Melrose  \& Watson 2006).
 Furthermore, the terms in intensity and intensity square expressions
  of Eq.(\ref{avin}) and Eq.(\ref{avi2n})  can be calculated by
 an integral of  the probability distribution
 over the 6 random variables (see Melrose  \& Watson  2005), for
 instance,
\bea &&
 <jk>  =  \la u_j u_k^* \ra  =  \int [ \int_{0}^{\infty}
d\phi_{j}'' \int_0^{\infty} d\phi_{k}''  + \int_{-\infty}^{0}
d\phi_{j}'' \int_{-\infty}^{0} d\phi_{k}''  +  \int_{0}^{\infty}
d\phi_{j}'' \int_{-\infty}^{0} d\phi_{k}'' \, e^{i \pi/2} \nn \\
&+& \int_{-\infty}^{0} d\phi_{j}'' \int_{0}^{\infty} d\phi_{k}'' \,
e^{-i\pi/2}]
\frac{P(\phi_{j},\phi_{k},\phi_{j}',\phi_{k}',\phi_{j}'',\phi_{k}''
)}{\left\vert 1 + r_{\rm F}^{2}\phi_{j}'' \right\vert^{1/2} \,
\left\vert 1 + r_{\rm F}^{2} \phi_{k}'' \right\vert^{1/2}}
\exp\left[i Z_{jk}\right] d\phi_{j} d\phi_{k} d\phi_{j}'
d\phi_{k}'\;, \eea with
 \be Z_{jk}  =  \rft ({\phi_j'}^{2} - {\phi_k'}^{2})/2
 + (\phi_{j} - \phi_{k})\;, \ee
 \be
 P(\phi_{j},\phi_{k},\phi_{j}',\phi_{k}',\phi_{j}'',\phi_{k}'')
 =P(\phi_{j})P(\phi_{k})P(\phi_{j}')P(\phi_{k}')P(\phi_{j}'')P(\phi_{k}'').
 \ee

\section{results and discussions}
\begin{figure}
\vspace{50mm}
   \begin{center}
\includegraphics{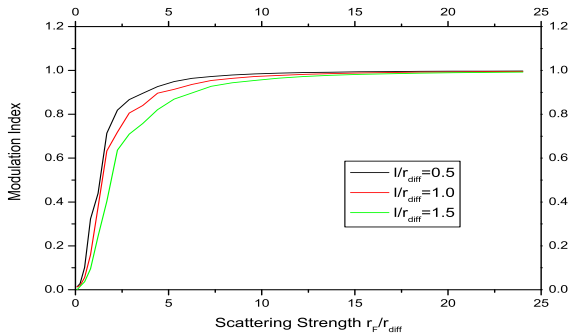}
%\vspace{60mm}
\caption{ Modulation index  v.s. scattering strength $\rf/\rdiff$.
The influence by the inner scale of ISM is plotted as indicated in
the figure.} \label{mod-af}
   \end{center}
   \end{figure}

The numerical results of the modulation index versus the scattering
strength $\rf/\rdiff$ are shown in figure \ref{mod-af}. We find that
 the modulation index increases with the scattering strength continually
  from  the weak scattering regime, via the transitional scattering
  regime, to the strong scattering regime where the modulation index
  is fully saturated. The influence by the inner scale on the
  modulation index is studied, and we find that the
  increasing of inner scale makes the
  modulation  index decrease a little  at the transitional
  scattering regime but little effect in the strong scattering regime.
  The trends of the modulation index in the weak and
   strong scattering regimes
  are similar to those obtained by the approximated treatments (see. e.g.
  Narayan 1992; Rickett 1977, 1990), however the SPP method provides
     the descriptions of modulation index at the transitional
     scattering.  If N=1,    a weak  scattering case, we have
\be m \approx  U_{1111}/U_{11}^{2} - 1\;, \ee from which we can
obtain that the modulation index increases with the scattering
strength $\rf/\rdiff$, similar to the case discussed by Salpeter
(1967) (Melrose \& Watson 2005). In the transitional scattering the
number N is about 2 or 3 but depends on the choice of inner scale.
If $N\rightarrow\infty$, a strong scattering case,
 we find that  the interchange terms of amplitude products, for
 instance $<jklm>$ with j,k,l and m being not all same,
 are very small and
$ U_{1111} \sim U_{1122} \sim U_{1221}) \sim 1/N^{2}$.
Therefore, the  following approximation  is preserved,
\be
m = <I^{2}>_{N} -1  \simeq
N U_{1111} +  N(N-1) (U_{1122} + U_{1221}) - 1 \simeq 1\;,
\ee
a fully modulated scintillation,
which has been obtained and discussed by Gwinn et al. (1998).
%\be mod^{2}(diff) \simeq N(N-1) (U_{1122} +
%U_{1221})/(N U_{11})^{2} - 1 = 1 \ee

{\bf Moreover, there exists a  discrepancy between our results and
those earlier results by Goodman and Narayan (2005) who found the
modulation index
 m$>$1 at transitional scattering regime, which
calls into question the reliability of the SPP method. Or it may
possibly be due to the 1-D nature of our  computation, which could
be resolved by applying the method to a 2-D screen in a future
work.}

%I think that the author should compare
%his work with their figure 2 and at least mention the differences.
%There are two possible considerations a) the SSP method is an
%approxiamtion that is not very accurate near the transition from weak
%to strong scattering; b) that the 1-D nature of the calculation causes
%the reduced m (I am doubtful of this - but I cannot be sure).

%\vskip 0.2cm
{\bf Acknowledgments} Thanks are due to many  helpful
  discussions  with   D.B. Melrose,  B. Rickett ,  D. Stinebring, M. Walker,  J.P. Macquart,
  N. Wang and X.J. Wu.
  %The author is very grateful for the critic comments  and discussions
   % from/with B. Rickett   that greatly improved the quality of the paper.

%\begin{thebibliography}{}
\section*{References}
\begin{description}
\itemsep 0mm

%%
%\item
%N. D. Ramesh Bhat, Yashwant Gupta, A. Pramesh Rao
%Astrophys.J. 500 (1998) 262-279
%Pulsar Scintillation and the Local Bubble

\item[]{} Armstrong~J.W., Rickett~B.J., Spangler~S.R. 1995, ApJ, 443,
209

%\item[]{} Cordes~J.M., Rickett~B.J., Stinebring~D.R., Coles~W.A.
%2004 (In preparation)

\item
Blandford, R.D. \& Narayan, R., 1985, MNRAS,  213, 591

\item[]{} Cordes~J.M., Weisberg~J.M., Boriakoff~V. 1985, ApJ, 288, 221

\item
Goodman, J.J., \&  Narayan, R. 1985, MNRAS, 214, 519
%\beta>4

\item
Goodman, J.J.,  \& Narayan, R. 2005, ApJ, 636, 510

\item
Gwinn, C.R. et al. 1998, ApJ, 505, 928
 %Interstellar Optics,

\item
Kaspi, V. M., \& Stinebring, D.R. 1992,  ApJ, 392, 530
%DR &quot;Long Term Pulsar Flux Monitoring and Refractive Interstellar
%     Scintillation,&quot;

%\item
%Kaplan D.L., Condon J.J., Arzoumanian Z., Cordes J.M., 1998, ApJS 119, 75

%\item
%M.L.A. Kouwenhoven, Astron. Astrophys. Suppl. Ser. 145, 243-254 , 2000
%Pulsars in the Westerbork Northern Sky Survey
%0.4 refractive scin, e-mail: M.L.A. Kouwenhoven@astro.uu.nl

%\item Berry, M.V., Diffractals, J. Phys. A: Math. Gen., {\bf 12}(6), 781 (1979)

%\item
%Buckley, R.,
%1971, Aust. J. Phys., 24, 351
%Diffraction by a random phase screen with very large
%R.M.S. phase deviation. I. One-dimensional phase screen,

%\item
%Buckley, R.,
%1971, Aust. J. Phys., 24, 373
%Diffraction by a random phase screen with very large
%R.M.S. phase deviation. I. two-dimensional phase screen,

%\item Grimmett, G.R. \& Stirzaker, D.R., Probability and Random Processes,
%Clarendon Press, Oxford, 1992

\item
Lyne, A.G., \& Smith, F.G. 2005,  Pulsar Astronomy,  Cambridge
University Press

\item
Manchester, R.N., Taylor, J.H. 1977, Pulsars, Freeman, San Fransisco
%DISS

\item
Mandel, L. \& Wolf, E. 1995, Optical Coherence and Quantum Optics,
Cambridge University Press

%\item Mercier, R.P. 1962, Proc. Camb. Phil. Soc., A58, 382

\item
Melrose, D.B., \& Watson, P.  2006, ApJ, 647, 1131

\item
Narayan, R. 1992, Phil. Trans. R. Soc. Lond. A,   341, 151
%physics of pulsar scintillation,

%\item
%Nye, J.F. 1999, Natural Focusing and Fine Structure of Light, IOP
%Publishing Ltd

\item Rickett, B.J. 1977, ARAA 15, 479

\item
 Rickett, B.J.  et al. 1984, A\&A 134, 390

\item
Rickett, B.J. 1990, ARAA 28, 561

\item
Rickett, B.J. 2006, ChJAA, 6, Suppl 2.

\item
Salpeter, E.E., 1967,  ApJ, 147, 433
% Interplanetary scintillations.
%\item  Salpeter, E.E. 1969, Nat 221, 31
%DISS

\item
Scheuer, P.A.G. 1968, Nat 218, 920
%DISS

\item
  Stinebring, D.J., et al. 2000,
  % T. V. Smirnova, T. H. Hankins, J. S. Hovis,
  %V. M. Kaspi, J. C. Kempner, E. Myers, and
   %   D. J. Nice
      ApJ, 539, 300
%Five Years of Pulsar Flux Density Monitoring:
%Refractive Scintillation and the Interstellar
%      Medium

%\item
%Stinebring, D.R>, Faison M.D., McKinnon, M.M.  1995,
%Refractive and Diffractive Scintillation of the Pulsar PSR~B0329+54
%studied by

\bibitem[]{}
Walker~M.A., Wardle~M.J. 1998,  ApJL, 498, L125

\item
Walker, M.A.,   Melrose, D.B.,  Stinebring, D.R., \& Zhang, C.M.
   2004,  MNRAS,  354, 43
%Interpretation of parabolic arcs in pulsar secondary

%\item Wolfram, S.,  The Mathematica Book,
%Cambridge University Press 1996

%\item
%Wang, N., Wu, X. J., \& Manchester, R. N., et al. 2001, ChJAA, 1, 430
% refractive scin

%\item
%Wang, N,  Manchester, R. N.,  Yusup, A., et al.
% Wu, X.J.,  Zhang, J.,  Chen, M. Z.
 %2001, A\&SS. 278, 57
% scintillation Observations of Strong Northern Pulsars

\item
 Wang, N.,  Manchester, R. N., Johnston, S.,, Rickett, B., Zhang, J., Yusup, A., \&  Chen, M.
 2005, MNRAS, 358, 270
% Long-term scintillation observations of five pulsars at 1540 MHz

\item
Watson, P. G., \& Melrose, D. B., 2006, ApJ, 647, 1142

%\item
% Zhou, A.Z.,  Wu, X.J., \& Esamdin, A. 2003,  A\&A,  403, 1059
 %-1065 (2003) % scintillation

\end{description}
%\end{thebibliography}
%\newpage

\end{document}